\documentclass[pdflatex,sn-mathphys]{sn-jnl}

\usepackage{amsmath}
\usepackage{multirow}

\usepackage{subfigure}
\usepackage[justification=centering]{caption}
\newcommand{\tabincell}[2]{\begin{tabular}{@{}#1@{}}#2\end{tabular}}

\jyear{2021}%

\theoremstyle{thmstyleone}%
\theoremstyle{thmstyletwo}%

\theoremstyle{thmstylethree}%

\begin{document}

\title[CNN-based Classification Framework with Auxiliary Information]{CNN-based Classification Framework for Lung Tissues with Auxiliary Information}

\author[1]{\fnm{Huafeng} \sur{Hu}}\email{huafeng.hu@xjtlu.edu.cn}

\author[2]{\fnm{Ruijie} \sur{Ye}}\email{sgrye2@liverpool.ac.uk}

\author*[3]{\fnm{Jeyarajan} \sur{Thiyagalingam}}\email{t.jeyan@stfc.ac.uk}

\author*[2]{\fnm{Frans} \sur{Coenen}}\email{Coenen@liverpool.ac.uk}

\author*[4]{\fnm{Jionglong} \sur{Su}}\email{jionglong.su@xjtlu.edu.cn}

\affil[1]{\orgdiv{Department of Electrical and Electronic Engineering}, \orgname{University of Liverpool based at Xi'an Jiaotong-Liverpool University}, \orgaddress{\city{Suzhou}, \postcode{215123}, \state{Jiangsu}, \country{China}}}

\affil*[2]{\orgdiv{Department of Computer Science}, \orgname{University of Liverpool}, \orgaddress{\city{Liverpool}, \postcode{L69 3BX}, \country{United Kingdom}}}

\affil*[3]{\orgdiv{Scientific Computing Department}, \orgname{ Science and Technologies Facilities Council}, \orgaddress{\street{Harwell Campus}, \city{Oxford}, \postcode{OX11 0GD}, \country{United Kingdom}}}

\affil*[4]{\orgdiv{School of AI and Advanced Computing}, \orgname{Xi'an Jiaotong-Liverpool University}, \orgaddress{\city{Suzhou}, \postcode{215123}, \state{Jiangsu}, \country{China}}}

\abstract{Interstitial lung diseases are a large group of heterogeneous diseases characterized by different degrees of alveolitis and pulmonary fibrosis. Accurately diagnosing these diseases has significant guiding value for formulating treatment plans. Although previous
work has produced impressive results in classifying interstitial lung diseases, there is still room for improving the accuracy of these techniques, mainly to enhance automated decision-making. In order to improve the classification precision, our study proposes a convolutional neural networks-based framework with auxiliary information. Firstly, ILD images are added with their medical information by re-scaling the original image in Hounsfield Units. Secondly, a modified CNN model is used to produce a vector of classification probability for each tissue. Thirdly, location information of the input image, consisting of the occurrence frequencies of different diseases in the CT scans on certain locations, is used to calculate a location weight vector. Finally, the Hadamard product between two vectors is used to produce a decision vector for the prediction. Compared to the state-of-the-art methods, the results using a publicly available ILD database show the potential of predicting these using different auxiliary information.}

\keywords{Deep Learning, Multimodal Data Learning, Image Classification, ILD Analysis}

\maketitle

\section{Introduction}\label{sec:1}

Interstitial lung disease (ILD) represents a group of more than 200 types of chronic inflammation that cause the scarring of human lung tissue, often affecting the lung parenchyma, the small lung airways, and the alveoli \cite{kreuter2015exploring}. The use of Computed Tomography (CT) lung scans is generally considered to be the most appropriate mechanisms for ILD diagnosis \cite{ingegnoli2012interstitial}. However, ILDs are difficult to distinguish, even by experienced physicians. They are a histologically heterogeneous group of diseases, and their clinical manifestations are similar. The result is that some ILDs are misdiagnosed due to subjectivity on behalf of radiologists inspecting lung scans and consequently a 
low diagnostic accuracy is typically encountered \cite{sluimer2006computer}. The idea of the research is to apply the tools and techniques of machine learning to aid and support ILD classification \cite{depeursinge2012near}. 

A number of approaches founded on machine learning have been used for ILD classification. Early approaches were founded on traditional machine learning methods such as: $k$-NN \cite{huber2010classification}, SVM \cite{lim2011regional} and random forest \cite{huber2012texture}. However, because of the limitations imposed by the processing power available at that time, these early approaches operated using a limited number of feature (sometimes extracted by hand). More modern deep learning methods, particularly CNNs, are able to process entire images. CNNs can automatically and efficiently idenrify relevant features from images, and have been shown to produce excellent results in the context of medical image analysis \cite{krizhevsky2012imagenet}. Deep neural networks, such as AlexNet \cite{krizhevsky2012imagenet}, VGGNet \cite{simonyan2014very}, GoogLeNet \cite{szegedy2015going}, ResNet\cite{he2016deep}, and Xception \cite{chollet2017xception} have all been shown to produce competitive results. However, these deep learning-based approaches tend to focus on the image data, auxiliary data is typically not taken into consideration. Medical images, such a CT lung scans, contain useful auxiliary information, such as the contrast between bones and organs. It is argued in this paper that the utilisation of this auxiliary information can improve the effectiveness of computer-aided approaches directed at ILD. This paper investigates the use of location information as auxiliary information. Location information consists of the occurrence frequencies of different ILDs in CT scans at different locations. However, to the best of our knowledge, in others' papers for ILDs classification, the location information is not used as a given knowledge. To the best of our knowledge, we are the first to use location information to give more accurate results for ILD classification. 

Given the foregoing, in this paper, we propose a novel CNN-based framework incorporating
auxiliary location information for ILD classification. An additional novel aspect is the use of the Housefield unit for recasting CT three channel lung scans. The Hounsfield unit is a dimensionless value, used in computed tomography, that allows for the differentiation between organs and tissues \cite{depeursinge2015optimized,gao2018holistic}. Next, we combine two deep CNNs for feature extraction. Finally, we incorporate the location information with the output from the previous model to improve the classification performance. The novelty of our research is twofold. First, we design a new input data form to enhance the features. Second, we employ the location information by the association between location and the lung tissues. It is used as another type of auxiliary information to increase lung disease classification performance. The main contributions are summarized as follows.

\begin{itemize}

	\item [1.] We design three new window ranges for ILD dataset to correspond to the three dimensions of the color map using the Hounsfield unit, which could be used as additional medical information to improve the classification performance of our framework; 
	
	\item [2.] We novelty demonstrate that there is a significant statistical relationship between location and disease through statistical analysis. Therefore, to the best of our knowledge, we are the first to employ location information as auxiliary information to improve the classification performance for ILD; and
	
	\item [3.] We combine Xception and Inception-v3 to extract the features of different tissues and obtain a decision vector by a CNN-based model incorporating both medical and location information. Compared to state-of-the-art methods, it outperforms at least $2.1\%$ for classifying the lung tissues, which is more suitable for analyzing ILD. 
	
\end{itemize}

The rest of this paper is organized as follows. Section~\ref{sec:2} gives a brief  literature review of relevant research. Section~\ref{sec:3} provides a detailed account of the evaluation data set used and the proposed approach. Section~\ref{sec:4} reports on the evaluation of the proposed approach and the results obtained. Section~\ref{sec:5} concludes the paper and gives some indicators for future research.

\section{Related Work}
\label{sec:2}

\subsection{Medical Image Classification with CNNs}

CNN may be traced back to Fukushima et al.'s proposal around 1988 \cite{fukushima1988neocognitron}. In 1989, LeCun et al. used backpropagation to train a CNN to classify patterns of handwritten numbers \cite{lecun1989backpropagation}. CNN was used in various applications in the early 1990s, including object recognition, character recognition, and face recognition. In 1995, Lo et al. were the first to apply CNN to medical image analysis, training a CNN to detect lung nodules in chest radiography \cite{lo1995artificial}. In 1997, Chan et al. used CNN to detect microcalcification on mammograms, and in the following year, they used it to detect the mass \cite{chan1997computerized,chan2005computer}. In 1994, Hasegawa et al. employed a distributed parallel CNN to recognize anatomical features on digital chest radiographs \cite{hasegawa1994convolution}. Although the early CNN is not enough for in-depth design to improve accuracy and applicability due to design and machine calculation problems, the potential of pattern identification power of CNN in medical pictures was proved.

CNN-based deep neural systems are successfully implemented for medical classification field. Because CNN is a suitable feature extractor, identifying medical images can avoid the time-consuming process. Anthimopoulos et al. presented a customized CNN with LeakyReLU to classify image patches of lung disease \cite{anthimopoulos2016lung}. Furthermore, they later discovered that utilizing restricted data makes it challenging to build an adequate model, as well as a CNN-based system could be learned from the original design and fine tuned using data from lung tissue \cite{christodoulidis2016multisource}. As a result, CNN transfer learning is widely employed in medical image analysis. Kermany et al. used transfer learning to train a neural network using a fraction of the data required by traditional methods \cite{kermany2018identifying}. Most of the experiments recevied suitable sensitivity but low specificity, whereas the CNN-based method had good sensitivity and specificity. Furthermore, the researchers also tested their technique on a small pneumonia dataset. Finally, this technique may aid in a faster diagnosis, allowing for early treatment and a higher cure rate. Minaee et al. also looked at using transfer learning to create a medical image classification system, an essential part of a CNN model with transfer learning \cite{minaee2020deep}. 

\subsection{ILD Classification with CNNs}

Since the 2010s, there has been much research, and significant progress, directed at using digital image processing and pattern recognition techniques for ILD identification. Early work adopted traditional machine learning methods is to extract the representations from the raw data for further classification. Deep learning techniques, particularly CNN, have recently been shown to be effective in the field of computer vision. Compared to the traditional method of manually selecting features, CNNs can autonomously learn more intrinsic and representative new features from the training data. In addition, the successful CNN model is more general, and it is easier to transfer design solutions to different types of problems, which makes it more widely concerned.

There has been significant work at using machine learning for ILD classification. Van et al. modified a restricted Boltzmann machine and used it to learn features for classifying lung tissues, 
incorporating sharing weights by the convolutional block \cite{van2014learning}. Li et al. designed a shallow CNN to classify lung image patches \cite{li2014medical}. Due to the insufficient number of data sets available, Anthimopoulos et al. proposed CNNs with LeakyReLU activations for radiologists. Moreover, In \cite{gao2018holistic}, a pre-trained deep CNN, AlexNet, was used to classify ILD scans after fine tuning the CNN with additional lung CT data. The required input to the AlexNet was color images with a size of $224\times224$, so the input images were resized and extended to three channels by a different range of windows with a medical quantitative scale called the Hounsfield unit. However, according to a suitable range for different organs on the Hounsfield unit, the selection range of three channels could be further adjusted. Several representation deep networks, such as VGGNet and ResNet, have be shown to be effective in various computer vision applications. Shin et al. discussed their descriptive ability for lung tissue categorization and obtained relatively good results by transfer learning \cite{shin2017three}. However, using these CNN models without modification, the classification performance of ILD is not as perfect as other image classification tasks, so researchers began to modify the model to make it more suitable for the characteristics of the ILD dataset. 

Advanced CNN models that are too deep have surpassed the information that the ILD dataset can carry, and the overfitting has become another problem that needs to be overcome \cite{tajbakhsh2016convolutional}. To address the problem of limited training data, in \cite{o2017comparison}, the authors demonstrated that modifying the CNN architecture could improve the classification performance. Huang et al. presented a novel two-stage transfer learning technique that applies information learned from source data and supplemental unlabeled lung data to improve the performance of ILD classification \cite{huang2020deep}. However, these networks overcome the overfitting problem by reducing the depth of the model rather than the data itself, thus limiting its best performance. Therefore, overcoming these problems becomes the key to breaking through existing CNN frameworks.

\subsection{Classification with Location Information}

Location information plays an important role in accurate classification in object detection and classification domains. In \cite{ding2018prior}, an indoor object recognition framework was proposed by employing location knowledge. In \cite{tang2015improving}, a successful CNN model incorporating the global position system coordinates was designed for image classification. In the context of biomedical applications, \cite{ghafoorian2017location} incorporated location information concerning White Matter Hyperintensities (lesions in the brain that show up in MRI scans). Furthermore, in the public ILD dataset at University Hospitals of Geneva \cite{depeursinge2012building}, the locations of the ROIs are also given. As noted earlier, in previous research, location information has not been appropriately used for ILD classification.

\section{Proposed Approach}
\label{sec:3}

\begin{figure}[ht!]
	\centering
	\includegraphics[width=\linewidth]{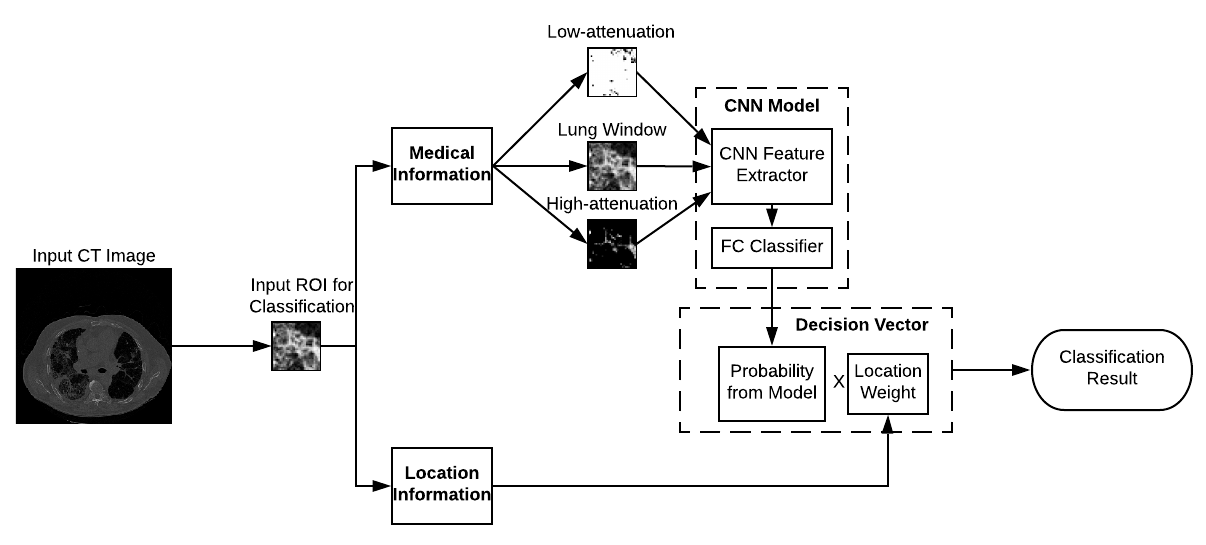}
	\caption{Our framework.}
	\label{fig:1}
\end{figure}

This section describes a novel CNN-based framework incorporating both medical and location information, used for ILD pattern classification. The framework, given in Figure~\ref{fig:1}, is constructed as follows. First, when a new lung disease classification request is received, the input image is appended with its medical information by extending to a three-channel input in the Hounsfield unit with their medical definitions. Specifically, three different ranges are used to improve the contrast between different tissues in lung CT. Second, a modified CNN model is used to produce a vector of classification probability for each lung tissue. Third, location information of the input image, consisting of the occurrence frequencies of different diseases in the CT scans on certain locations, is used to calculate a location weight vector. Fourth, the Hadamard product between two vectors produces a decision vector for the prediction. We normalize the sum of each class in the decision vector to 1. Finally, the classification result is the tissue with the maximum value in the decision vector. The details of the medical information, the CNN model, and the location information are given below.

\subsection{Medical Information}

The Hounsfield unit is a dimensionless unit universally used by radiologists in the interpretation of CT images \cite{razi2014relationship}. It's calculated by transforming the baseline linear attenuation coefficient into a linear function, where water is defined to be 0 Hounsfield unit and air defined as -1000 Hounsfield unit \cite{hounsfield1980computed}. Because their different ranges can correspond to different organs and regions, Hounsfield units are used to evaluate and quantify tissues, and fluids \cite{motley2001hounsfield}. In recent years, it also has been used for determining the appropriate model of treatment \cite{gucuk2014usefulness}. Therefore, we use the Hounsfield unit to extend our input data with medical representations. 

According to the transformation from pixel value to Hounsfield unit, the range of values for our dataset is from -1000 to 1800. Furthermore, the regions with over 400 Hounsfield unit are simply bones with different radiodensity \cite{razi2014relationship}. Therefore, a commonly used range of thresholds in the lung disease classification is [-1000, 400]. 

\begin{figure} [ht]
	\centering
	\subfigure[Orignial.]{
		\includegraphics[height=1in]{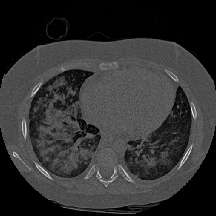}
		\label{fig:2a}
	}
	
	\subfigure[Low attenuation.]{
		\includegraphics[height=1in]{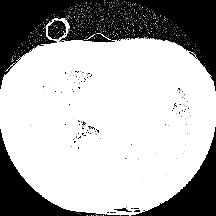}
		\label{fig:2b}
	}\hspace{5mm}
	\subfigure[Lung window.]{
		\includegraphics[height=1in]{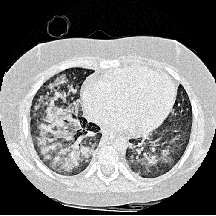}
		\label{fig:2c}
	}\hspace{5mm}
	\subfigure[High attenuation.]{
		\includegraphics[height=1in]{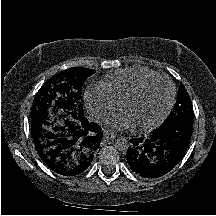}
		\label{fig:2d}
	}

	\caption{An illustration of an original lung CT slice. We extend the original image into three channels called low attenuation range, lung window range, and high attenuation range.}
	\label{fig:2}
	
\end{figure} 

Using a CNN architecture designed for color images, the value of the Hounsfield unit could be used as additional medical information to improve the classification performance of ILDs \cite{depeursinge2015optimized,gao2018holistic}. However, the previous research did not give a reasonable explanation for the selection of its range, and the accuracy of each range also needs to be further improved. Therefore, we expand all gray CT slice images into three CT windows with more suitable ranges: lung window range $[-1000, -50]$, high attenuation range $[-100, 400]$, and low attenuation range $[-1000, -700]$. Specifically, the low attenuation range is used to describe patterns with lower intensities, such as improving the differences between EM and other tissues; the lung window range is used to show normal regions of lung, and the high attenuation range is used to represent patterns with higher intensities. We extend the transformed images into three channels from the original input images. Meanwhile, instead of copying it on the three channels, the different CT windows help visualize certain texture patterns of lung diseases \cite{gao2018holistic}. The three-channel input with the values of the Hounsfield unit is used as the input into our model. Figure~\ref{fig:2} gives an example of the low attenuation range, the lung window range, and the high attenuation range from an original lung CT slice. Using three attenuation ranges can provide different image features between five lung tissues. Moreover, we adapt our CNN architecture from ImageNet with three channels using these three ranges.

\subsection{CNN}

\begin{figure}
	\centering
	\includegraphics[width=\linewidth]{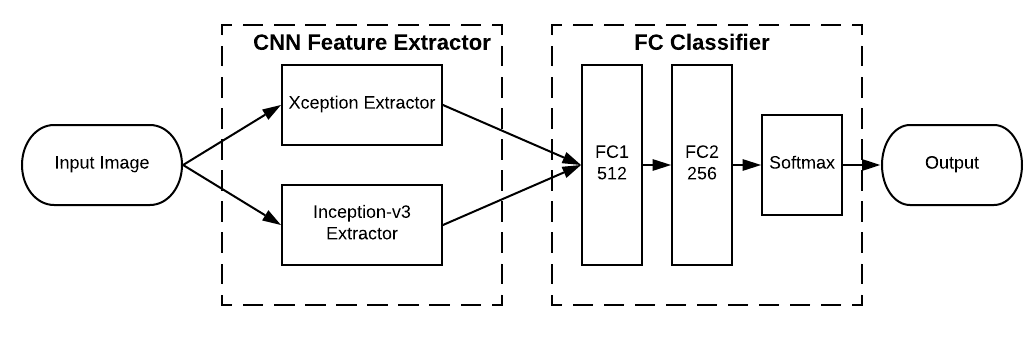}
	\caption{The CNN-based model during our research.}
	\label{fig:3}
	
\end{figure}

Although the CNN application achieves significant results on large data sets, on small sets, the performance of CNN is not enough for real world application \cite{shin2016deep}. In order to achieve better classification performance and reduce the computational costs, transfer learning is used to pre-train a model from the original domain to decrease the need for abundant labeled data for training \cite{shin2016deep}. Therefore, we choose two powerful CNN architectures, Xception \cite{chollet2017xception} and Inception-v3 \cite{szegedy2016rethinking}, with transfer learning from ImageNet as our initial parameters for the architecture \cite{shin2016deep}. In our experiments, our network contains multiple blocks, features of ILDs from Xception and Inception-v3, followed by two FC layers and a 5 classes softmax layer. The schematic architecture of our model is given in Figure~\ref{fig:3}. Let $p_c, c=1,2,3,4,5$ be the output from our CNN model for each class $c$. The vector $P$, such that $ P = (p_1,p_2,p_3,p_4,p_5)$, is used as a probability vector to predict the classification results.

\subsection{Location Information}

The locations of ROI are aptly described as apical, basal, diffuse, perihilar, peripheral/subpleural, and non-relevant \cite{depeursinge2012building}. Generally, the locations of ROIs are relevant to several diseases, e.g., the tissues of MN are usually found on the location called diffuse \cite{depeursinge2012building}. Therefore, we also use the information from location to produce a location weight vector to improve the classification performance of our model. Let $L_i,\ i=1,2,...,5$ be location weight vector of $i$-th location in our dataset as $L_i = (L_{i1},L_{i2},L_{i3},L_{i4},L_{i5})$, where $L_{ic}$ is the location weight of the $c$-th class in the location $i$. It could be calculated as,

\begin{equation}
    L_{ic} = \frac{\textrm{total}\ \textrm{number}\ \textrm{of}\ \textrm{class}\ c\ \textrm{on}\ \textrm{location}\ i}{\textrm{total}\ \textrm{number}\ \textrm{of}\ \textrm{samples}\ \textrm{on}\ \textrm{location}\ i}
\end{equation}
This vector is used as a location vector for the prediction.

\subsection{Decision Vector}

When a new image is fed into our CNN, it produces a vector of probability $P$ with additional medical information, which is obtained by three different attenuations with Hounsfield units. After that, the location information is employed according to the relationship between location and disease. The location information is used to improve the performance of our result by generating location weight vector $L_i$. The Hadamard product between two vectors is carried out to produce a decision vector. The sum of each class for each input is mapped to 1. The output prediction lung tissues of input data on location $i$ is the tissue class with the maximum value on the decision vector as,

\begin{equation}
\textrm{prediction}\ \textrm{class} = \textrm{arg}\mathop{\textrm{max}}_{c} \frac{P\circ L_{i}}{P\cdot L_i^T}
\end{equation}

where $P\circ L_{i}$ is the Hadamard product \cite{horn1990hadamard} between $P$ and $L_i$ and $L_i^T$ is the transpose of $L_i$.

\section{Results and Discussions}
\label{sec:4}

In this section, first, we depict the dataset used in this chapter. Next, we describe the experimental settings, the association between location and the lung tissues, as well as the evaluation and the discussion of our results.

\subsection{Dataset}

\begin{figure}[ht]
	\subfigure[NM]{
		\includegraphics[height=0.8in]{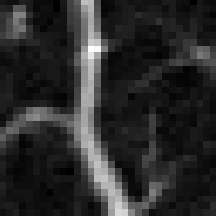}
		\label{fig:4a}
	}
	\subfigure[EM]{
		\includegraphics[height=0.8in]{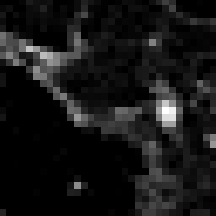}
		\label{fig:4b}
	}
	\subfigure[GG]{
		\includegraphics[height=0.8in]{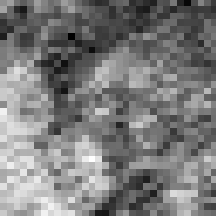}
		\label{fig:4c}
	}
	\subfigure[FB]{
		\includegraphics[height=0.8in]{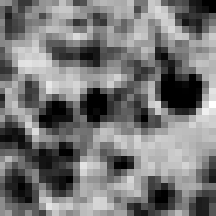}
		\label{fig:4d}
	}
	\subfigure[MN]{
		\includegraphics[height=0.8in]{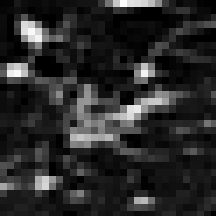}
		\label{fig:4e}
	}
	\caption{The five selected lung tissues for our work are NM, EM, GG, FB, and MN.}
	\label{fig:4}
\end{figure} 

In this chapter, we use a public database of ILDs containing 113 sets of CT scans \cite{depeursinge2012building}. Each slice contains $512\times 512$ pixels with ground truth annotations for 17 different lung tissues. Four of ILDs are the most common tissues in ILD patients and have been studied in the past \cite{depeursinge2012near,song2013feature,song2015large,li2014medical,gao2018holistic}: emphysema (EM), ground glass (GG), fibrosis (FB), 
micronodules (MN), and the normal tissue (NM) is also added. Therefore, our
method is used to classify five lung tissues. An illustration of their respective appearances is provided in Figure~\ref{fig:4}.
The images are divided into patches of $32\times 32$ pixels with at most $16\times 32$ overlap. In order to enhance the representativeness of the data, we exclude all patches where more than $25\%$ of the pixels are outside of these five lung tissues from the experiment. Thus, a total of 16531 image patches are used in our experiment. The numbers of
patches of each tissue class of lung together with the percentage of datasets are given in Table~\ref{tab:1}.

\begin{table}[h]
	\begin{center}
	\caption{Summary of the dataset used.}\label{tab:1}
	\begin{tabular}{c|c|c}
	\hline
		Class&Number of Patches& Percentage\\
		\hline
		NM &5289&$32.0\%$\\
		EM &1044&$6.3\%$\\
		GG &1955&$11.8\%$\\
		FB &2564&$15.5\%$\\
		MN &5679&$34.4\%$\\
		\hline
		
	\end{tabular}
	\end{center}

\end{table}

\subsection{Experimental Setup}

The experiments in our research are based on 5-fold cross-validation. Furthermore, all patches belonging to the same patient are included in the same fold. The experiment is thus performed on a total of 16531 image patches. We use four folds for training and hyperparameters fine tuning, meanwhile, test the rest for each cross-validation procedure. After that, we augment the image patches in the training set by rotations at degrees 90, 180, and 270 and horizontal and vertical flipping. We resize the input data into the most suitable resolution for different CNN models, e.g., $299\times299$ for Xception and Inception-v3, and $224\times224$ for AlexNet and VGG16. In the rest of this subsection, we describe the evaluation metrics we used and the implementation details.

\subsubsection{Evaluation Metrics}

In our experiments, two metrics are used for evaluating the performance of different methods. First, we use the $F$-score for the different classes because of the high sensitivity of it to imbalanced classes \cite{goutte2005probabilistic}. Second, the overall accuracy is also computed to evaluate the overall performance. 

The $F$-score is the harmonic mean of the precision and recall, the function is used as \cite{goutte2005probabilistic}
\begin{equation}
F\textrm{-score}_c= \frac{2\times \textrm{precision}_c\times \textrm{recall}_c}{\textrm{recall}_c+\textrm{precision}_c} 
\end{equation}
where $c=1,2,3,4,5 $ and

\begin{equation}
    \textrm{recall}_c=\frac{\textrm{samples}\ \textrm{correctly}\ \textrm{classified}\ \textrm{as}\ c}{\textrm{samples}\ \textrm{of}\ \textrm{class}\ c} 
\end{equation}

\begin{equation}
    \textrm{precision}_c=\frac{\textrm{samples}\ \textrm{correctly}\ \textrm{classified}\ \textrm{as}\ c}{\textrm{samples}\ \textrm{classified}\ \textrm{as}\ c}
\end{equation}

\begin{equation}
    \textrm{accuracy}=\frac{\textrm{correctly}\ \textrm{classified}\ \textrm{samples}}{\textrm{total}\ \textrm{number}\ \textrm{of}\ \textrm{samples}} 
\end{equation}

\subsubsection{Implementation}

The proposed method was implemented in Python with Keras framework \cite{geron2019hands}. All experiments were conducted on a computer with an Intel Core i7-6850K CPU of 3.60GHz, a GPU of NVIDIA GeForce GTX 1080 Ti, and 64GB of
RAM.

\subsection{Association between Location and the Lung Tissues}

\begin{figure}[ht]
	\centering
	\subfigure[Whole Dataset]{
		\includegraphics[height=1.5in]{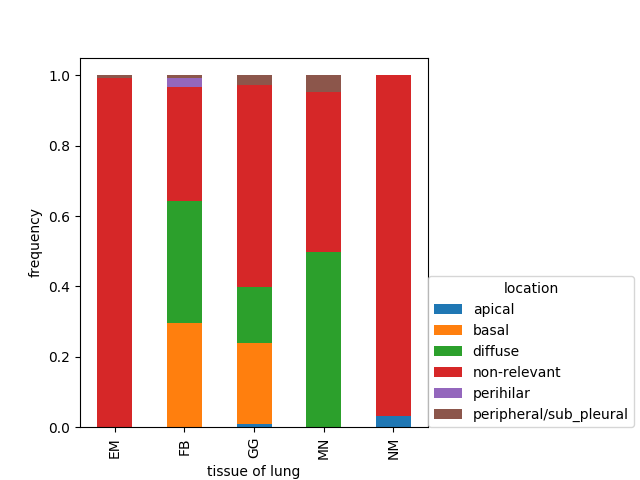}
		\label{fig:5a}
	}
	\subfigure[Dataset without Non-relevant]{
		\includegraphics[height=1.5in]{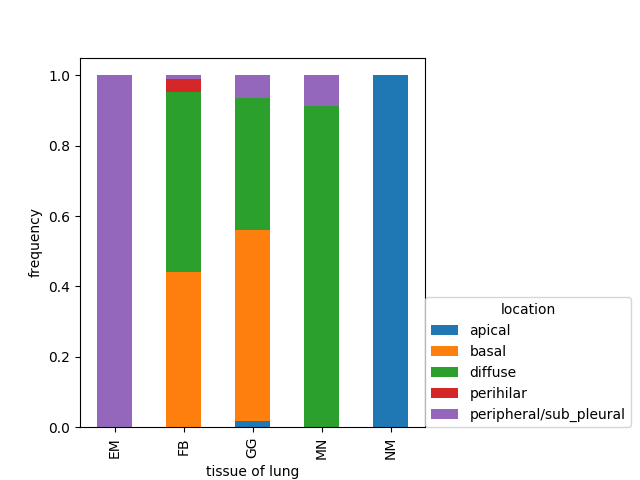}
		\label{fig:5b}
	}

	\caption{Two bar charts for our dataset.}
	\label{fig:5}
\end{figure}

In our dataset, we find five different lung tissues and six different locations of ROIs, i.e., apical, basal, diffuse, perihilar, peripheral$/$subpleural and non-relevant \cite{depeursinge2012building}. In definition, non-relevant means no prevailing location. Moreover, we perform to delete the influence of non-relevant and obtain two bar charts for whole images and images without non-relevant in Figure~\ref{fig:5}. It shows that without a non-relevant class, a certain lung tissue in a certain location may produce higher weights than the whole dataset.

To determine the association between locations and diseases, we apply different tests, namely, Pearson Chi-squared test with its corresponding $p$-value, and Cramer Chi-squared test. In statistics, Cramer Chi-squared is used to measure the association between two different variables \cite{acock1979measure}. The null hypothesis, $H_0$, for our Pearson Chi-squared test is that the location is independent of the lung tissues. 
\begin{table*}[htbp]
	\centering
	\caption{Statistical tests for correlation between locations and diseases.}
	\begin{tabular}{c|c|c}
	\hline
		& Whole Dataset & Dataset without Non-relevant\\
		\hline
		Pearson Chi-squared test & 874.75 & 455.45 \\
		
		$p$-value & $<2.2e-16$ & $<2.2e-16$ \\
		Cramer Chi-squared test & 0.3100434& 0.4952374 \\
		\hline
	\end{tabular}%
	\label{tab:addlabel}%
\end{table*}%

The statistical results are shown in Table~\ref{tab:addlabel}. The $p$-value for the Person Chi-squared test is smaller than 0.05. Thus, we reject the null hypothesis and conclude that there is an association between the location and the lung tissues. The value of the Crammer Chi-squared test shows that the relationship between the location and the lung tissues in the dataset without non-relevant is more significant than the whole dataset. Our data suggests a significant association between the location and the lung tissues, and we use a dataset without non-relevant for our location weight vector.

\subsection{Results}

\begin{sidewaystable}
	\centering
	\caption{Classification performance of our CNN with Different Settings.
	}
	\begin{tabular}{c|c|c|c|c|c|c}
		\hline
		Pooling Type& Pooling Percentages &Dropout Rate& FC1&FC2&$F$-score& Accuracy \\
		\hline
		Max&$100\%$&0.5&512&256&0.8627&$86.6\%$\\
	Average&$50\%$&0.5&512&256&0.8402&$84.1\%$\\
	Average&$25\%$&0.5&512&256&0.8173&$81.6\%$\\
	\hline
	Average&$100\%$&0.2&512&256&0.8866&$88.9\%$\\
	Average&$100\%$&0.3&512&256&0.8912&$90.0\%$\\
	Average&$100\%$&0.4&512&256&0.8896&$88.8\%$\\
	Average&$100\%$&0&512&256&0.8412&$84.3\%$\\
	
	\hline
	
	Average&$100\%$&0.5&512&512&0.8994&$90.1\%$\\
	Average&$100\%$&0.5&256&256&0.8961&$89.7\%$\\
	Average&$100\%$&0.5&128&128&0.8842&$88.3\%$\\
	Average&$100\%$&0.5&128&256&0.8869&$88.9\%$\\
	Average&$100\%$&0.5&256&128&0.8943&$88.5\%$\\

	Average&$100\%$&0.5&512&256&$\textbf{0.9044}$&$\textbf{90.8\%}$\\
		
		\hline

	\end{tabular}

	\label{tab:6}
	
\end{sidewaystable}

In this subsection, we discuss the experimental results in three parts, i.e., fine tuning of hyperarameters, the effect of using auxiliary information, and comparsion with the state-of-the-art. The following results and the selection of hyperparameters on the best resulting models with 5-fold cross-validation. This subsection is divided into three parts. First, we present a set of experiments that are used to choose components and fine tune the
hyperparameters. Second, we discuss the effect of medical and location information in our experiments. Third, we compare our method with state-of-the-art works under the same conditions. 

\subsubsection{Fine Tuning of Hyperparameters}

Here we demonstrate the choices of components and the fine tuning of hyperparameters for our method. The input of these results is the patches with the medical information. Table~\ref{tab:6} demonstrates the classification performance for different configurations of the network's architecture. Using the average pooling layer increases the classification performance by roughly $3\%$ compared to the max pooling layer. It means that in experiments, the average pooling layer may be more effective than the max pooling layer. Meanwhile, decreasing the pooling percentages from $100\%$ to $50\%$ or $25\%$, resulted in a decline of more than $6\%$ and $9\%$, respectively. 

To identify the dropout rate for our framework, we conduct experiments to investigate the effect of using a convolutional layer with different dropout rates. We chose a dropout rate of 0.2, 0.3, 0.4, and 0.5. The classification result obtains the best performance with a dropout rate of 0.5. When the dropout layers are removed, there is a drop in of more than $6\%$ because of overfitting. Furthermore, we change the number of neurons in FC1 and FC2. FC1 with 512 neurons and FC2 with 256 neurons achieve the best performance.

Table~\ref{tab:7} gives the performances of using different optimizers and loss functions for our model. We use SGD with a learning rate of 0.001 and a momentum of 0.90, Adam with 0.001 learning rate, and 0.0001 for AdaGrad. The loss function called categorical cross-entropy by the SGD optimizer produced the best results. The result of Adam is about $1\%$ lower performance and AdaGrad with a higher decrease of $3\%$. After that, we also produce a comparable result by using SGD with the loss function of MSE.

Finally, we use the hyperparameters that give the best performance. They consist of an average pooling layer with $100\%$, a dropout rate of 0.5, SGD with 0.0001 learning rate, and 0.9 momentum. Moreover, the first FC layer has 512 neurons and the second FC layer has 256 neurons.

\begin{table}[htb!]
	\centering
	\caption{Performance of the proposed CNN with different components and hyperparameters.
	}
	\begin{tabular}{c|c|c|c}
		\hline
	Optimizer& Loss Function & $F$-score& Overall Accuracy \\
		\hline
		Adam& Cross-entropy&0.8902&$89.1\%$\\
		
		AdaGrad	&Cross-entropy&0.8711&$87.0\%$\\
		
		SGD&MSE&0.8873&$88.8\%$\\
		SGD&Cross-entropy&$\mathbf{0.9044}$&$\textbf{90.8\%}$ \\

		\hline

	\end{tabular}

	\label{tab:7}
	
\end{table}

\subsubsection{The Effect of using Auxiliary Information}

Information contained in the different pretrained CNN architectures may be complementary, even when pretrained on the same data \cite{guerin2018improving}. Therefore, we combined the features from Xception and Inception-v3 for the input image independently, followed by the same FC layers and softmax layer for classification. Specifically, the simple CNN model uses the image intensity values within the window [ -1000, 400] in the Hounsfield unit and expand one channel into three channels. All the methods were implemented by 5-fold cross-validation. The location information is multiplied by the result from the CNN. For convenience, Model 1 represents the CNN-based framework that only incorporates medical information; Model 2 represents the CNN-based framework that incorporates both medical and location information. The confusion metrics of Model 1 and Model 2 are shown in Figure~\ref{fig:4}. Moreover, the comparison of framework without any information, Model 1, and Model 2 is summarized in Table~\ref{tab:4}. The increase in the classification performance for GG is around $5\%$. In comparison, the increase for EM, FB, and MN is around $1\%$. Results show that employing location information may improve the classification performance of our experiments. The results show that adding the medical information by the different intensities of ILDs outperformed the other models by $5\%$.

\begin{table*}[htb!]
	\centering
	\caption{Comparison of framework without any information, Model 1, and Model 2. Model 1 represents the CNN-based framework only incorporating medical information; Model 2 represents the CNN-based framework incorporating both medical and location information.
	}
	\begin{tabular}{c|c|c}
		\hline
		Methods& Average F-score& Overall Accuracy \\
		\hline
		Framework without any information& 0.8423&$84.5\%$\\
		
Model 1	&0.9044&$90.8\%$\\
		
		Model 2& $\mathbf{0.9215}$&$\mathbf{92.3\%}$\\

		\hline
		
	\end{tabular}

	\label{tab:4}
	
\end{table*}

\begin{figure*}[ht]
	\centering
	\subfigure[Model 1.]{
		\includegraphics[height=2in]{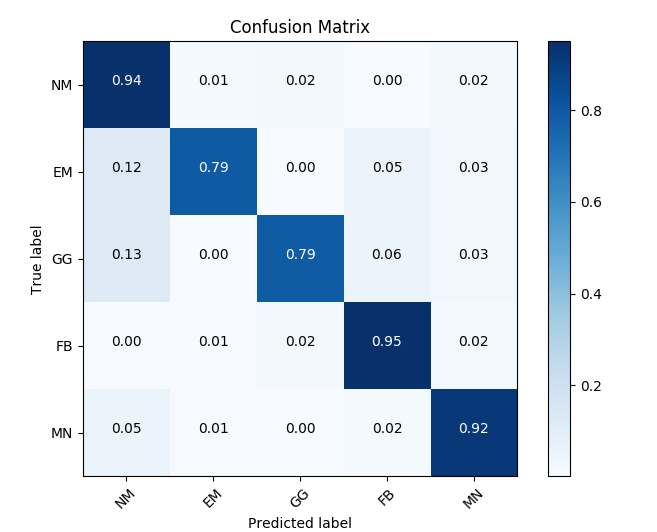}
		\label{fig:6a}
	}
	\subfigure[Model 2.]{
		\includegraphics[height=2in]{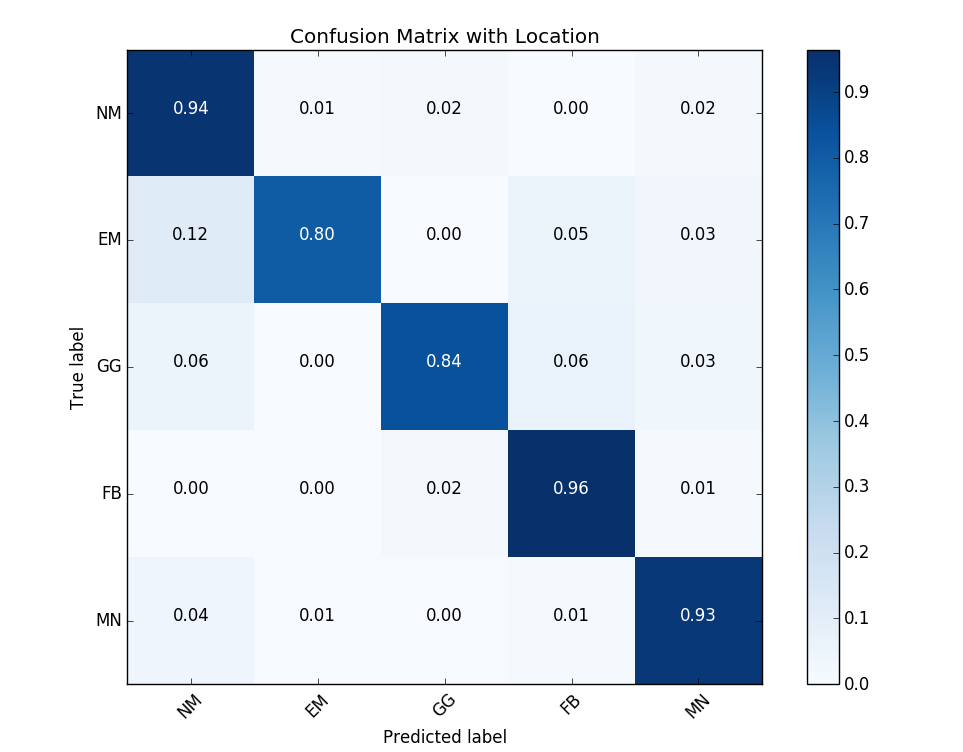}
		\label{fig:6b}
	}

	\caption{Confusion matrix of our dataset. Model 1 represents the CNN-based framework only incorporating medical information; Model 2 represents the CNN-based framework incorporating both medical and location information.}
	
		\label{fig:6}
\end{figure*}

Table~\ref{tab:13} gives some examples of classification results with the probabilities on top-3 classes. The tissue classes in bold are the classification results with the highest probabilities. (F$\|$T), as an example, Model 1 results in a false classification while our Model 2 results in correct classification. Similarly, we obtain (F$\|$F) and (T$\|$F).

Misclassification is unavoidable in cases where some lung patches of lung are challenging to classify, e.g., MN in Table~\ref{tab:13}. However, according to the experimental results, the misclassification ratio could be reduced by employing location information. The examples of GG and EM are correctly classified by our Model 2. The classification accuracy of Model 1 is around $90.8\%$ and, by employing location information, Model 2 achieves an additional $1.5\%$.

\begin{sidewaystable}
	\centering
	\caption{Classification results with their probabilities for top-3 classes. Model 1 represents the CNN-based framework only incorporating medical information; Model 2 represents the CNN-based framework incorporating both medical and location information.
	}
	\begin{tabular}{c|c|c|c|c|c}
	\hline
		Input Patch&True Class& Location&Model 1 &Model 2&Comparison \\
		\hline
		
		\parbox[c]{1em}{
			\includegraphics[width=0.3in]{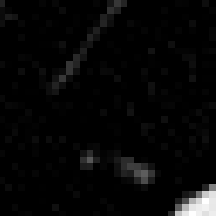}} & EM&Peripheral or subpleural & \tabincell{c}{(\textbf{NM}, EM, FB),\\ (\textbf{0.58},0.35,0.06)}&\tabincell{c}{(\textbf{EM}, NM, FB),\\ (\textbf{0.72},0.20,0.02)} &(F$\|$T)\\
		\hline
		\parbox[c]{1em}{
			\includegraphics[width=0.3in]{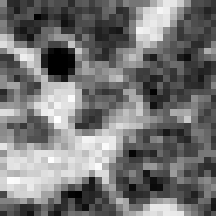}} & GG&Basal& \tabincell{c}{(\textbf{MN}, GG, FB),\\(\textbf{0.75},0.20,0.04)}&\tabincell{c}{(\textbf{GG}, EM, FB),\\(\textbf{0.71},0.25,0.03)} &(F$\|$T)\\
		\hline
		
		\parbox[c]{1em}{
			\includegraphics[width=0.3in]{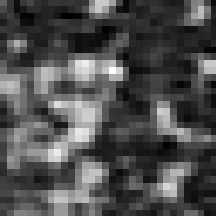}} & MN&Diffuse& \tabincell{c}{(\textbf{MN}, GG, FB),\\(\textbf{0.40},0.37,0.20)}&\tabincell{c}{(\textbf{GG}, MN, FB),\\(\textbf{0.41},0.36,0.21)} &(T$\|$F)\\
		\hline
		
		\parbox[c]{1em}{
			\includegraphics[width=0.3in]{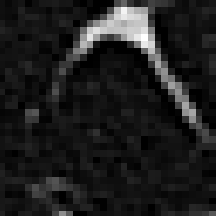}} & NM&Non-relevant & \tabincell{c}{(\textbf{GG},NM, EM),\\ (\textbf{0.64},0.28,0.08)}&\tabincell{c}{(\textbf{GG}, NM, EM),\\ (\textbf{0.64},0.28,0.08)} &(F$\|$F)\\
		
		\hline
	\end{tabular}

	\label{tab:13}
	
\end{sidewaystable}

\subsubsection{Comparison with the State-of-the-art}

We use two metrics for fair comparisons with state-of-the-art methods,i.e., $F$-score and overall accuracy, on the experiments. A comparison of the proposed framework with state-of-the-art methods by $F$-score is given in Table~\ref{tab:48}. In most classes, our proposed framework earns the highest $F$-scores.

\begin{table*}[htb!]
	\centering
	\caption{$F$-score for classifying ILDs by different method.
	}
	\begin{tabular}{c|c|c|c|c|c}
		\hline
		Methods& NM&EM& GG&FB& MN \\
		\hline
		Song \cite{song2013feature}&0.8400&0.7530&0.7820&0.8410&0.8570\\
		Song \cite{song2015large} &0.8770&0.7680&0.7950&0.8770&0.8880 \\
		
		Li \cite{li2014medical}&0.8395&0.5449&0.7150&0.7624&0.9096\\
		
		Gao \cite{gao2017holistic} &0.8844&$\mathbf{0.89}$&0.82&0.85&0.90\\
		
		Proposed Method&$\mathbf{0.9268}$&0.8312&$\mathbf{0.8688}$&$\mathbf{0.9340}$&$\mathbf{0.9524}$\\
		\hline

	\end{tabular}

	\label{tab:48}
	
\end{table*}

Table~\ref{tab:5} provides a comparison of our framework with other state-of-the-art methods on the classification accuracy level. We also evaluated the performance of the AlexNet \cite{krizhevsky2012imagenet}, and VGG16 \cite{simonyan2014very}, we resized the $32\times 32$ patches to $224\times 224$ and extended 3 channels by three different Hounsfield unit windows as same as before to suit the input requirement of these two networks. Furthermore, we also combined these two models after fine tuning the model for evaluation. The combination network outperformed single AlexNet and single VGG16 by $4\%$ and $2\%$. However, the classification performance is still inferior to our proposed. The results demonstrate the superior performance of our scheme to others, outperforming the rest by $2\%$ to
$15\%$. In conclusion, both the $F$-score and overall accuracy show that the location and medical information could improve the classification performance.

\begin{table*}[htb!]
	\centering
	\caption{Overall accuracy of ILD classifications.
	}
	\begin{tabular}{c|c}
		\hline
		Methods& Overall Accuracy \\
		\hline
		Li \cite{li2014medical}& $67.1\%$\\
		
		Song \cite{song2015large} &$86.1\%$\\
		
		Shin \cite{shin2016deep} &$90.2\%$\\
		
		Anthimopoulos \cite{anthimopoulos2016lung} & $85.6\%$\\
		
		Joyseeree \cite{joyseeree2018rotation} &$80.3\%$ \\
		
		AlexNet & $83.4\%$\\
		
		VGG16 &$85.2\%$\\
		AlexNet + VGG16 & $87.8\%$\\
		
		Proposed Method&$\textbf{92.3\%}$\\
		
		\hline
		
	\end{tabular}

	\label{tab:5}
	
\end{table*}

\section{Conclusions and Future Work}
\label{sec:5}

In this paper, we present a CNN-based framework incorporating both medical and location information that improves
the accuracy and stability of ILDs classification. In our design, medical and location information was employed to improve the classification performance of ILDs. The experiments demonstrate that the additional medical and location information enhances the framework for ILDs classification. Moreover, two sets of different features from Xception and Inception-v3 are useful for improving classification accuracy. The proposed framework outperform the state-of-the-art methods about $ 2\%$-$15\%$ in classification accuracy. The reported performances could obtain the best classification result, it proves the potential of employing medical and location information that could be beneficial for classifying ILDs. Moreover, we believe that other medical datasets with auxiliary information may also improve the classification results. 

For the ILDs dataset, each patient has more than 10 slices, in our experiments, we just used the patches of image as the input. Therefore, the information between different slices from the same patient is not rationally used in our framework. Our future research on the topic is to use a 3D dataset instead of a 2D dataset. The plans are twofold. First, the relationship between different slices from the same patient could be used as another information for improving the result of ILDs classification or other medical image classification. The aim of this future research is to combine the information from the same patient to improve the classification performance. The possible plan is to use a time series-like model to combine the information. Second, while deep CNNs have succeeded in 2D medical image analysis, analyzing essential tissues or structures from 3D medical images remains a tough challenge for CNNs. The aim of this future research is to propose a 3D CNN for end-to-end medical image analysis. The simple way is to change the dimensions of input data or connect a number of 2D CNN for the classification.

\bibliography{huafeng_hu_CNN.bib}

\end{document}